\documentclass[prb,showpacs,twocolumn,floatfix]{revtex4}
\usepackage[english]{babel}
\usepackage{color}
\usepackage{graphicx}
\usepackage{amsmath}
\usepackage{amssymb}
\usepackage{epsfig}

\renewcommand{\vec}[1]{\mathbf{#1}}
\newcommand{\veck}{{\mathbf{k}}}

\newcommand{\Tr}{{\rm Tr}}
\newcommand{\bsigma}{\boldsymbol{\sigma}}

\begin{document}
\title{Topologically stable gapless phases of time-reversal invariant superconductors}
\author{B. B{\'e}ri}
\affiliation{Instituut-Lorentz, Universiteit Leiden, P.O. Box 9506,
2300 RA Leiden, The Netherlands}
\date{September 2009}
\begin{abstract}
We show that time-reversal invariant superconductors in $d=2$ ($d=3$) dimensions can
support topologically stable Fermi points (lines), characterized
by an integer topological charge. Combining this with the momentum space
symmetries present, we prove analogs of the fermion doubling
theorem: for $d=2$ lattice models admitting a spin$\times$electron-hole structure, 
the number of Fermi points is a multiple
of four, while for $d=3$, Fermi lines come in pairs. 
We show two implications of our findings for topological superconductors in $d=3$: 
first, we relate the bulk topological invariant to a topological
number for the surface Fermi points in the form of an index theorem. 
Second, we show that the existence of topologically stable Fermi lines results
in extended gapless regions in a generic topological superconductor phase diagram.
\end{abstract}
\pacs{74.20.-z,74.25.Jb,73.20.At} 
\maketitle{}
\section{Introduction}
Time-reversal ($\cal T$) invariant superconductors have attracted a lot of
attention lately as they provide  possibility to realize topologically
nontrivial gapped phases of matter.\cite{schnyderclass,schnyderAIP,kitaevtable,qi:187001,roy2008topological,schnydermodel,ryumodel,chZTI}
While topologically nontrivial $\cal T$-invariant gapped phases of $^3$He were discussed
already back in the eighties\cite{cosmic,volovik1989fractional,volovik1992exotic} (these phases also appear in 
the recent constructions,\cite{qi:187001,roy2008topological} see also Ref.~\onlinecite{volovik2009}),
the key observation highlighting the general role of  ${\cal T}$ and
electron-hole (EH) symmetries in the topological classification of gapped \mbox{${\cal T}$-invariant} superconductors
 has been made only recently: Schnyder {\it et al}\cite{schnyderclass} has noted
that combining ${\cal T}$ and EH symmetries, one finds that the Bogoliubov
de Gennes Hamiltonian
\begin{equation}
{\cal H}_{BdG}(\veck)=
\begin{pmatrix}
H(\veck)&\Delta(\veck)\\
\Delta^\dagger(\veck) & -H^*(-\veck)
\end{pmatrix}
\label{eq:HBdG}\end{equation}
obeys a chiral symmetry: ${\cal C}{\cal H}_{BdG}{\cal C}^{-1}=-{\cal
  H}_{BdG}$, with  ${\cal C}={\cal C}^\dagger$ unitary. In Eq.~\eqref{eq:HBdG},  the $N\times N$ matrices $H(\veck)$ and $\Delta(\veck)$ are the single particle Hamiltonian and the
pairing potential. There exists a basis\cite{schnyderclass} in
which the chirality operator is ${\cal C}'={\rm   diag}(\openone_N,-\openone_N)$, 
and the Hamiltonian
takes the block offdiagonal form, 
\begin{equation}
{\cal H}'_{BdG}(\veck)=
\begin{pmatrix}
0 & D(\veck)\\
D^\dagger(\veck)& 0
\end{pmatrix}.
\label{eq:offdiag}\end{equation}
A gapped system,  as the determinant of $D({\bf k})$
is nonvanishing for all $\bf k$, realizes a mapping from the momentum space to 
the space $GL(N,\mathbb{C})$ of invertible $N\times N$ matrices. 
In $d=3$, such mappings fall into distinct homotopy classes, which directly translates
to the possibility for topologically nontrivial three dimensional gapped systems, i.e., systems which are not deformable
into a $\bf k$-independent gapped  model without a gap closing somewhere throughout the deformation.\cite{TREHfootnote}
Such topologically nontrivial gapped superconductors are called topological superconductors. 
(Topological superconductors also exist in \mbox{$d=1,2$},\cite{schnyderclass,schnyderAIP,kitaevtable,qi:187001,roy2008topological,volovik1989fractional}
but the ones considered in this work are three dimensional.)
The homotopy class of the system is characterized by an integer bulk topological invariant
$\nu$,\cite{schnyderclass} constructed such that its nonzero value signals a topological superconductor.

Besides their value in analyzing  gapped
phases, homotopy arguments prove to be useful in studying systems
without a  gap.\cite{Volovikbook,Horava,Volovikqpt} In such systems, 
homotopy arguments can be used
to determine the possible dimensionalities of  topologically stable
Fermi surfaces, defined as manifolds of zero energy that cannot be destroyed by a
small deformation of the Hamiltonian. 
A Fermi surface is stable if a nontrivial
topological charge can be assigned to it; the possible values of the topological charge
also result from the homotopy considerations.

As we have illustrated, the block-offdiagonal structure in Eq.~\eqref{eq:offdiag} provides key insights
into the topology of gapped phases of $\cal T$-invariant superconductors. In this work we would like
to study what insights it could provide regarding the nature of  gapless
phases of these systems.
One motivation for  this, naturally, lies in the characterization of the possible nodal
structures of systems such as high-$T_c$ materials, or noncentrosymmetric superconductors.\cite{Sato}
The reason, however, why this question is especially timely is given by the relevance of gapless
systems in the context of topological superconductors: first, as $\nu$ does not change under
non-gap closing deformations of the theory, the system must become gapless across a topological
phase transition\cite{Volovikbook,Volovikqpt} (a transition between gapped phases with different $\nu$).
Second, the surface of a topological superconductor can also be thought of as
a two dimensional gapless system: it is known to support 
topologically protected gapless modes,\cite{schnyderclass,schnyderAIP} robust against arbitrary
deformations (provided that these do not close the bulk gap).

With these motivations in mind, our first objective is to present a homotopy argument leading
to the topological characterization of Fermi surfaces of $d=2,3$ dimensional \mbox{$\cal T$-invariant} superconductors. 
This is carried out in Sec.~\ref{sec:FSandq}. In Sec.~\ref{sec:2d} we deduce the consequences of our findings for two dimensional systems, including
the surface of a topological superconductor. The consequences in three dimensions
are discussed in Sec.~\ref{sec:3d}, together with the implications regarding topological
phase transitions. Our conclusions are summarized in Sec.~\ref{sec:concl}.

\section{Fermi surfaces and topological charge} 
\label{sec:FSandq}
Due to $\det {\cal H}'_{BdG}(\veck)=-|\det D({\bf k})|^2$, the vanishing of the gap at
a  point ${\bf k}_0$ on a Fermi surface is in  one-to-one correspondence with the vanishing
 of the  determinant of $D({\bf k})$ at this
point. We first assume that ${\bf k}_0$ is not a time-reversal invariant momentum (TRIM),
i.e., there is no such reciprocal lattice vector ${\bf G}$ that ${\bf k}_0= -{\bf
  k}_0+{\bf G}$. Then,  in the vicinity of ${\bf k}_0$, 
the matrix $D({\bf k})$ need 
not satisfy any restrictions; all
restrictions stemming from EH  and ${\cal T}$
symmetries relate\cite{schnyderclass} $D({\bf k})$ to $D({-\bf k})$. 
Therefore, in the vicinity of ${\bf k}_0$ the only restriction on ${\cal   H}'_{BdG}(\veck)$
 is chiral symmetry. The properties of Fermi surfaces in $d=2$ 
 systems  with this restriction were studied by Wen and
 Zee.\cite{WenZee} Their considerations can be formulated in a compact way
 which also includes the $d=3$ case as follows  
(see also Refs.~\onlinecite{Volovikbook,Horava,Volovikqpt}):
for a Fermi surface  of dimension $d-p$, in the $p$ dimensions transverse to
 it,  $\det D({\bf k})\neq 0$ in the vicinity of the surface. Then,
embracing the point ${\bf k}_0$ on
the Fermi surface with a small $p-1$ sphere $S^{p-1}$ in the transverse dimensions, we have a mapping from
 $S^{p-1}$ to the
space $GL(N,\mathbb{C})$. If this
represents  a nontrivial class in the homotopy group
$\pi_{p-1}(GL(N,\mathbb{C}))$, there is no small deformation of $D({\bf  k})$ 
which would make the map contractible to the inside of $S^{p-1}$, thus would 
remove the point of $\det D({\bf k})=0$ within $S^{p-1}$, i.e., the Fermi surface is stable. For $p=1,2,3$ only for
$p=2$ is there a nontrivial homotopy group,
$\pi_{1}(GL(N,\mathbb{C}))=\mathbb{Z}$. This leads to our central result: 
${\cal T}$-invariant superconductors can support $d-2$ dimensional topologically stable Fermi surfaces, which can
be assigned a topological charge taking values in the additive group of
integers. 

Fermi surfaces in $d=3$ ($d=2$) are thus Fermi lines (Fermi points). 
Their dimensionality can be also understood by noting that
 a Fermi surface is a manifold of $\det D({\bf k})= 0$,
i.e., it is defined by two real equations for $\veck$.\cite{ddimnote} 
The topological charge can be found by noting that the sphere in the transverse space is
 $S^{1}$, i.e., it is a circle. The topological invariant $q$ characterizing mappings from the circle to
$GL(N,\mathbb{C})$ is simply given by the winding number of the phase of $\det D({\bf k})$. 
The topological invariance of this charge means that $q$ does not change under deformations of the circle into any
loop around the Fermi surface, nor under deformations of $D({\bf   k})$, provided that the
loop does not cross a Fermi surface during these changes. Following Ref.~\onlinecite{WenZee}
(see also Refs.~\onlinecite{Volovikqpt,Manes}), we can express this integer charge directly through the Hamiltonian as
\begin{equation}
q=\frac{-1}{4\pi i}\oint \Tr\left[{\cal C}'({\cal H}'_{BdG})^{-1}\nabla_{l}{\cal H}'_{BdG}\right]d{l},
\label{eq:vortex}\end{equation}
where the integral is for a loop around the Fermi surface. 

Due to its dimensionality, and due to the fact that $q$ is the winding of the phase, a Fermi surface can be viewed as 
a momentum space vortex in $\det D({\bf   k})$ as an "order parameter".
Henceforth, we will often use this vortex picture; we emphasize that in this work, 
vortices, antivortices always refer to these momentum space objects and not to real space defects in the
superconducting order parameter.
Under a deformation of 
$D({\bf k})$,  local\cite{localnote} creation of Fermi surfaces  proceeds
through formation of vortex loops in three dimensions, and vortex-antivortex
pairs in two dimensions. Processes in the opposite direction are also possible:
a large enough deformation can shrink and remove vortex loops, or annihilate
vortex-antivortex pairs; this explains why the notion of topological stability
provides robustness only against small perturbations.
For lattice systems, instead of referring to the
way the Fermi surfaces are created, a more general statement can be
formulated, similarly to Ref.~\onlinecite{NiNino}: assume that $D(\veck)$ is smooth and
periodic in momentum space. In $d=2$, the net winding is then zero in the reciprocal unit cell
 and is equal to  the sum of the topological charges of the Fermi points. 
In $d=3$, a similar statement can be made for two dimensional sections
of the reciprocal unit cell parallel to its faces, with the intersections
of Fermi lines with a section playing the role of Fermi points.

At this stage, it is worthwhile to compare our result to earlier
findings for single band systems.\cite{Hatsugai,Volovikqpt,Sato} While these
works also find $d-2$ dimensional Fermi surfaces, Ref.~\onlinecite{Sato}
 finds a topological charge with $\mathbb{Z}_2$ values instead of $\mathbb{Z}$
values. (Ref.~\onlinecite{Volovikqpt} found a $\mathbb{Z}$ charge for a
spin-degenerate, single band system.)  For comparison, we consider the
two-dimensional  example of high-$T_c$ materials from
Ref.~\onlinecite{Sato}. The single particle Hamiltonian and the pairing
potential are
\begin{eqnarray}
H(\veck)&=&\alpha(\veck^2-k_F^2)+\vec{g}(\veck)\bsigma,\\
\Delta(\veck)&=&i\Delta_0(k_1^2-k_2^2)\sigma_2+i\vec{d}(\veck)\bsigma,
\label{eq:highTc}
\end{eqnarray}
where $\sigma_j$ are the Pauli matrices in spin-space, and the band parameter $\alpha$, 
the singlet pairing strength $\Delta_0$, and the Fermi momentum $k_F$ are
positive constants. The vectors  $\vec{g}(\veck)$ and
$\vec{d}(\veck)$, corresponding to spin-orbit coupling and the triplet pairing strength, respectively, 
are real  odd functions of $\veck$.  For
$\vec{g}(\veck)=\vec{d}(\veck)=0$, there are four Fermi points at $\veck_0=(\pm k_{\rm F},\pm
k_{\rm F})/\sqrt{2}$, which split into pairs of Fermi points once $\vec{g}(\veck)$,
$\vec{d}(\veck)$ become nonzero. Ref.~\onlinecite{Sato} finds that the Fermi points
for $\vec{g}(\veck)=\vec{d}(\veck)=0$ have zero charge and they split up into pairs with
unit charge, in agreement with the $\mathbb{Z}_2$ sum rule. As a Fermi point with zero
charge is not topologically stable, an alternative scenario seems also
possible, in which the system becomes  gapped due to a small
deformation. Applying our considerations shows that this is not the case. 
After a suitable transformation diagonalizing the chirality operator, the
offdiagonal block of ${\cal H}'_{BdG}(\veck)$ becomes $D(\veck)=\Delta(\veck)+H(\veck)\sigma_2$.
For $\vec{g}(\veck)=\vec{d}(\veck)=0$, $\det D(\veck)=\xi(\veck)^2$
where the winding of the phase of $\xi(\veck)$ around each Fermi point is of unit magnitude, hence
for each Fermi point $|q|=2$. Under a small deformation, the
splitting into two Fermi points with $|q|=1$ is thus the generic behavior. 
The example here suggests Ref.~\onlinecite{Sato} to be consistent with our findings upon 
identifying the $\mathbb{Z}_2$ charge as \mbox{$q$ mod $2$.}

\section{Consequences in two dimensional models}
\label{sec:2d}
\subsection{Fermion quadrupling in lattice models}
\label{sec:quadrupling}
After discussing aspects for which the restrictions relating $D(\veck)$ to
$D(-\veck)$ are unimportant, we turn to study the implications of these restrictions.
Time-reversal invariant superconductors can be classified depending on
symmetries under spin-rotation: they are said to belong to class DIII if there is
no spin rotation symmetry, to class AIII if $\sigma_3$ is conserved and to
class CI if there is full spin-rotation symmetry.\cite{Alt97,heinzner2005symmetry,fosterludwig} The restrictions on
$D(\veck)$ are as follows\cite{schnyderclass}:
\begin{subequations}\label{eqs:Drel}
\begin{align}
D(\veck)&=-D^T(-\veck)\quad \textrm{(class DIII)},\\
D_\sigma(\veck)&=D_{\bar{\sigma}}^T(-\veck)\quad\textrm{(class AIII)},\label{eq:relAIII}\\
D_\sigma(\veck)&=D_{\bar{\sigma}}(\veck)=D_\sigma^T(-\veck)\quad
\textrm{(class CI)}.\label{eq:relCI}
\end{align} 
\end{subequations}
For class AIII, a chiral symmetry
\mbox{${\cal C}'={\rm   diag}(\openone_{N/2},-\openone_{N/2})$} applies for each spin separately.
From these relations we see that if ${\bf k}_0$ is on a Fermi surface, so
is $-{\bf k}_0$. Regarding the further consequences of Eqs.~\eqref{eqs:Drel}, 
we first focus on the two dimensional case, i.e., the Fermi surfaces are Fermi points. 
We consider the generic situation that Fermi points have $|q|=1$ (not counting spin
in class CI); a Fermi point with a
higher $|q|$ splits into Fermi points with $|q|=1$ upon a slight deformation of $D(\veck)$.
We start with  class DIII and CI. Let us first assume that ${\bf   k}_0$ is not a TRIM. 
Taking a loop around ${\bf   k}_0$, and the momentum reversed of this
loop around $-{\bf   k}_0$, calculating the phase winding of $\det  D(\veck)$ on these loops, 
we immediately find that $q$ of a Fermi point at ${\bf   k}_0$ and at $-{\bf
    k}_0$ is the same.
 What happens if ${\bf   k}_0$ is a TRIM? For class CI, for each
point on a circle centered at a TRIM, the determinant of $D({\bf k})$ at the
antipodal points of the circle agree, leading to an even winding number.
Thus, if there is a Fermi point at a TRIM, a deformation of $D({\bf k})$ splits it
into  Fermi points off the TRIM, leading us back to a case satisfying our original
assumption. The same reasoning can be applied for class DIII, provided that $D(\veck)$ is even
dimensional. 
This is guaranteed if ${\cal H}_{BdG}$ admits a spin$\times$EH
structure, since it implies that its dimension is a multiple of four.
For class
AIII, if there is a Fermi point at ${\bf k}_0$ for spin up, there is an
other at $-{\bf k}_0$ for spin down.

If the conditions for zero net winding hold, each vortex should have
its antivortex (for class AIII, for each spin separately). 
It follows from our preceding discussion that the momentum reversed pairs cannot be
also the antivortex pairs. Starting from a Fermi point at ${\bf k}_0$, we thus immediately obtain
three other. That is, the combination of Eqs.~\eqref{eqs:Drel} with our findings regarding 
the topology of Fermi points has led us
to a result analogous to the fermion doubling theorem:\cite{NiNino}
the number of Fermi points is a multiple of four;  for class CI each Fermi point is spin degenerate
(i.e., we have a multiple of eight including spin), for
class AIII, the number of Fermi points is a multiple of two for each spin.   
We note that Ref.~\onlinecite{schnydermodel} conjectures a similar result, 
stating that the number of Fermi points is an even multiple of the minimal number 
of Dirac cones in a given symmetry class. This agrees with our findings for classes AIII and CI, but it gives
the number of Fermi points to be  a multiple of two for class DIII. 
With our proof at hand we see that this is not possible in models with a spin$\times$EH structure.

\subsection{Surface states of topological superconductors}
A two dimensional surface separating topological superconductors with different bulk
topological invariants $\nu_+$ and $\nu_-$ is known to support  
topologically protected Fermi points.\cite{schnyderclass,schnyderAIP}
(The theory for the low energy excitations on the surface 
defines a two dimensional $\cal T$-invariant superconductor model, hence 
its gapless modes form Fermi points.) 
Considerations based on three dimensional Dirac models\cite{schnyderclass,schnyderAIP} suggest
the number of surface Fermi points
 to be $|\nu_+-\nu_-|$. 
This equality cannot hold in all cases: while $\nu_+-\nu_-$ is a topological invariant, i.e., it 
cannot change under deformations (which do not close the gap), the number of surface Fermi
points is not, as such deformations are allowed to add or remove vortex-antivortex pairs in the surface theory. 
Note, however, that similarly to the violation of the fermion doubling theorem on the surface of a topological insulator,\cite{fukanemele} the fermion quadrupling theorem can be violated
in the above domain wall by allowing the Fermi points to  have a  nonzero net
  winding number $q_{\rm net}$ (i.e., it is possible to have vortices without  antivortex pairs). 
This number, as it is quantized in integer values, also cannot change by
continuous deformations of the theory: it is also a topological invariant. 
(Stated in another way,  $q_{\rm net}$ measures the number of {\it unpaired} vortices; this clearly cannot change
by adding or removing vortex-antivortex {\it pairs}.)
It is tempting to look for an "index theorem" relating these topological invariants. 
For this relation, we can use the results from the Dirac models as a reference.
For a domain wall in the $x-y$ plane with
$\nu(z\rightarrow\pm\infty)=\nu_\pm$, the 
$|\nu_+-\nu_-|$ Fermi points of the surface theory are characterized\cite{schnyderclass} by
a $D(\veck)$ equivalent to $(k_1+\eta ik_2)\openone_{|\nu_+-\nu_-|}$, where $\eta={\rm sgn}(\nu_+-\nu_-)$. 
Importantly, all Fermi points are unpaired: they all have identical unitary winding $q={\rm sgn}(\nu_+-\nu_-)$.
 The net winding is thus
\begin{equation}
q_{\rm net}=\nu_+-\nu_-,
\label{eq:indexth}
\end{equation}
providing us with the index theorem we were looking for. 
This equation, due to the topological invariance of both sides, holds irrespective to the possible changes
in the total number of surface Fermi points. For this total number, assuming again that all Fermi points have
\mbox{$|q|=1$}, $|\nu_+-\nu_-|$ sets the minimum: once all the Fermi points are unpaired, their number cannot be 
further reduced by pair annihilation. This also immediately explains the topological protection of the
gapless surface states. 

As a special case, Eq.~\eqref{eq:indexth} includes the surface of a topological superconductor, 
i.e, an interface between a topological superconductor ($\nu=\nu_+$) and
the vacuum ($\nu_-=0$). In this case, the significance of our index theorem 
is that it gives a direct measure of the {\it bulk}
topological invariant of the superconductor in terms of a topological number for the {\it surface} degrees of freedom.

It is worthwhile to go through the possible values of $q_{\rm net}$ for the various symmetry
classes in Eq.~\eqref{eqs:Drel}. Our reasoning in Sec.~\ref{sec:quadrupling} shows that $q_{\rm net}$
is always even in class CI (not counting spin). 
This is in accord with the observation\cite{schnydermodel} that $\nu$
can take only even values in this symmetry class. For classes DIII and AIII (considering only one spin
in the latter case), $\nu$ can take any integer value. This is again in accord with our previous
considerations if we note that for class DIII, the surface theory model is not constrained to
have a spin$\times$EH structure. 

Finally, our index theorem can be also viewed as completing
an analogy between  (${\cal T}$-breaking) chiral superconductors in two dimensions, and ${\cal T}$-invariant
topological superconductors in three dimensions. Regarding bulk properties, the two classes of systems are analogous
in that both are characterized by a bulk topological invariant $\nu$ taking values in $\mathbb{Z}$.
For a domain wall in chiral superconductors, there is an index theorem\cite{Volovikbook} equating $\nu_+-\nu_-$ with
 the number of  unpaired edge states (i.e., the difference between the number of left moving and right moving
modes in the domain wall). Our index theorem with the number of unpaired Fermi points on its left hand side thus nicely 
completes the analogy for the boundary degrees of freedom.

\section{Consequences in three dimensions}
\label{sec:3d}

\subsection{Fermi lines come in pairs}
\label{sec:FLpairs}
In three dimensions, a $d-2$ dimensional Fermi surface is a Fermi line. 
We now prove our next result: the consequence of 
Eqs.~\eqref{eqs:Drel} is that Fermi lines come in
pairs. While for class AIII, this is simply due to a Fermi line
with spin up having a pair with spin down, classes CI and DIII deserve some
discussion. 
We consider a lattice system with reciprocal unit cell defined by
$\sum_i\alpha_i\vec{b}_i$, $\alpha_i\in (-1/2,1/2)$. (Here $\vec{b_i}$ are primitive reciprocal lattice
vectors). The half cell bounded by the plane $\cal P$ at $\alpha_3=0$ and the
face ${\cal P}'$ with $\alpha_3=1/2$ is mapped to the other half of the cell
upon $\veck\rightarrow -\veck$. Due to Eqs.~\eqref{eqs:Drel}, 
a Fermi line located entirely in one of the half cells (not intersecting $\cal
P$ nor $\cal P'$) has a pair in the other half cell. 
The situation is not so clear for Fermi lines intersecting ${\cal P}$ and/or
${\cal P'}$;  to prove our assertion we now show that in the generic situation
a Fermi line cannot be its own pair under $\veck\rightarrow -\veck$.
First note the following facts:  
the TRIM of our reciprocal unit cell are on ${\cal P}$, ${\cal P'}$.
The behavior of ${\cal P}$, ${\cal P'}$ under $\veck\rightarrow -\veck$
is that of the two dimensional case, thus the intersection points with Fermi lines
have the earlier discussed properties of 
Fermi points in two dimensions. In particular,
generically there are no intersection points at TRIM.
 Now assume that there is a Fermi line which is its own pair, which
intersects  ${\cal P}$ and ${\cal P'}$ in the set
of points $\{\veck_j\}\cup\{\veck'_j\}\cup\{-\veck_j\}\cup\{-\veck'_j\}$
($\veck_j\in{\cal P}$, $\veck'_j\in{\cal P'}$). Consider the pair of points
$\veck_1$, $-\veck_1$ with winding number $q$. Take two parametrizations of
the line, $\vec{a}(t)$ and $\vec{b}(t)$, $\vec{a}(0)=\veck_1$,
$\vec{b}(0)=-\veck_1$, $\vec{a}(t)=-\vec{b(t)}$. Since the Fermi line is a
vortex, for a given parametrization, Fermi points on $\cal P$, $\cal P'$ with
like charges are pierced from the same direction (always from above or always
from below). As $\veck_1$ and $-\veck_1$
have the same charges and since $\vec{b}$ pierces $-\veck_1$ from the opposite
direction to how $\vec{a}$ pierces $\veck_1$, $\vec{a}$ and $\vec{b}$
parametrize the line with opposite orientation. 
Thus the two parametrizations must run into each other at some $t_0$, where
$\vec{a}(t_0)=\vec{b}(t_0)+\vec{G}$, which implies that $\vec{a}(t_0)$ is a
TRIM, which does not happen in the generic situation. 
This thus proves that lines come in pairs also for classes DIII and CI. 
While for class CI, it would also possible, as for class AIII,  to define pairs in terms of
spin, as we saw, more can be said: there are spin degenerate pairs of lines.
Note that in our proof we have not used the criterion of zero net
winding. This however does not lead to further doubling of the number
of lines,  as it can be satisfied for example with pair of Fermi loops.

\subsection{Phase diagram of topological superconductors}
\begin{figure}[t]
\includegraphics[width=0.45\textwidth,clip]{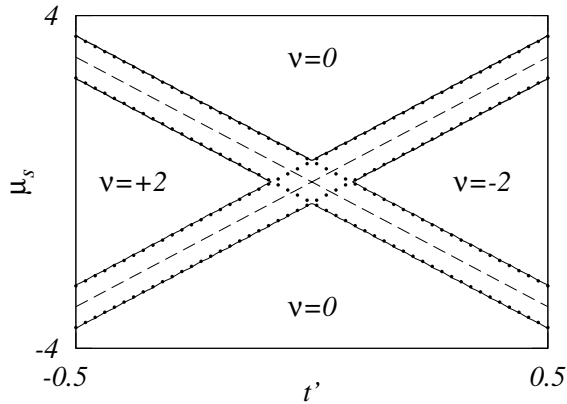} 
\caption{
\label{fig:phasediag}
The phase diagram of the model in Eq.~\eqref{eq:Dwmu}. The dashed lines are
the phase boundaries for $\mu=0$ [i.e., for the model in Eq.~\eqref{eq:D0}]
running at \mbox{$\mu_s=\pm 6t'$.} The solid lines are the numerically
obtained phase boundaries for $t=4$, $\Delta=2$, $\mu=0.5$; they enclose a gapless region in parameter space.
With dots, we also displayed $\mu_s=\pm 6t'\pm \mu$, defining the
phase boundaries according to Eq.~\eqref{eq:pbound}.}
\end{figure}

A topological phase transition, i.e., a transition between gapped phases with different bulk topological invariant $\nu$
proceeds through a gapless phase.\cite{Volovikbook,Volovikqpt} If the gapless phase can support topologically stable
Fermi surfaces, it is expected to be a stable region in
parameter space;\cite{murakami} small variations of the system parameters should not
lead out of it. 
In earlier models of topological superconductors,\cite{ryumodel,schnydermodel,chZTI}
however, topologically distinct gapped phases are directly adjacent to each other in parameter space; 
a slight deformation of a gapless system on the phase boundary between them immediately leads to one of the gapped
phases. The gapless phase is thus unstable. 
As a final application, we show how the existence of topologically stable Fermi lines can modify this situation. 
We consider the class CI
model of Schnyder {\it et al},\cite{schnydermodel} constructed for the diamond lattice. It is defined by
\begin{eqnarray} \label{eq:D0}
D(\veck)
=
\left(
\begin{array}{cc}
\Delta_\veck -{i}\Theta_\veck  & -{i}\Phi_\veck \\
-{i}\Phi^*_\veck & \Delta_\veck+{i}\Theta_\veck \\
\end{array}
\right) ,
\end{eqnarray}
with
\begin{align}\nonumber
\Phi_\veck
&=
2t 
\left[
e^{+{i} \frac{k_z}{4}}
\cos\frac{k_x+k_y}{4}
+
e^{-{i} \frac{k_z}{4}}
\cos\frac{k_x-k_y}{4}
\right],
\end{align}
\begin{eqnarray}\nonumber
\Delta_\veck
\!\!&=&\!\!
4 \Delta
\Big[
\cos \frac{k_x}{2} 
\cos \frac{k_y}{2}
-
\cos \frac{k_y}{2} 
\cos \frac{k_z}{2}
-
\cos \frac{k_z}{2} 
\cos \frac{k_x}{2}
\Big],
\nonumber
\end{eqnarray}
\begin{eqnarray}\nonumber
\Theta_\veck
\!\!&=&\!\!
4t'
\cos \frac{k_z}{2}
\left(
\cos \frac{k_y}{2}-\cos \frac{k_x}{2}
\right)
+\mu_s,
\end{eqnarray}
where the real numbers $t$, $t'$ parametrize the nearest neighbor and next-nearest neighbor hopping amplitudes, 
$\Delta$ is a real pairing strength, and  $\mu_s$ is a staggered potential. 
The phase diagram is shown in Fig.~\ref{fig:phasediag}. The system is gapless
on the lines $\mu_s=\pm 6 t'$, with the gap closing occurring at momenta 
$K_{1,\pm} = 2\pi \left(\begin{array}{ccc}\pm 1/3, & 1, & 0 \end{array}
\right),\,
K_{2,\pm} =
2\pi
\left(
\begin{array}{ccc}
1,& \pm 1/3, & 0
\end{array}
\right)$.
Gapless systems  belong to an unstable region in
parameter space: a slight change for example in $\mu_s$ immediately opens a gap. 
This can be traced back to the fact that the model satisfies
\begin{equation}
D(\veck)=\tau_2 D^*(\veck)\tau_2, 
\label{eq:extrasymm}
\end{equation}
where $\tau_j$ are Pauli matrices in the grading of Eq.~\eqref{eq:D0}. 
Consequently,  $\det D(\veck)=\Delta_\veck^2+|\Phi_\veck|^2+\Theta_\veck^2$, thus the space 
of invertible $D(\veck)$ matrices is  $\mathbb{R}^4\setminus\{0\}$. As \mbox{$\pi_{p-1}(\mathbb{R}^4\setminus\{0\})=0$}
($p=1,2,3$), the model is indeed not expected to have stable gapless phases. 
With Eq.~\eqref{eq:extrasymm} in addition to Eq.~\eqref{eq:relCI},  the model has more symmetries than required for belonging to class CI (i.e., it is "not generic"). The symmetry  \eqref{eq:extrasymm} can be removed by introducing a nonzero chemical potential $\mu$.  (The $\mu=0$ case corresponds to half-filling.) This leads to 
\begin{equation}
D'(\veck)=D(\veck)+i\mu \openone_2, 
\label{eq:Dwmu}\end{equation}
which, due to $\det
D(\veck)=(\Delta_\veck+i\mu)^2+|\Phi_\veck|^2+\Theta_\veck^2$, modifies the
space of invertible matrices.
To see how the nature of the gapless phase is
modified, we assume that $t'$, $\mu_s$, $\mu$ is small compared to $t$,
$\Delta$, and 
expand $D'(\veck)$
around the points $K_{j,\pm}$, resulting in
\begin{equation}
D'_{j,\pm}(\vec{p})=(p_1+i\mu)\openone_2+i\tau_1p_2+i\tau_2p_3-i\tau_3\tilde{m}_\pm,
\end{equation}
where $\vec{p}$ is a scaled momentum parametrizing the deviation from  $K_{j,\pm}$, and $\tilde{m}_\pm=\mu_s\pm6t'$. 
The condition for a gapless phase is now 
\begin{equation}\label{eq:pbound}
\mu^2-\tilde{m}_\pm^2=p_2^2+p_3^2, \quad p_1=0,
\end{equation}
describing the appearance of Fermi lines through the transition. In this more generic, 
$\mu\neq 0$ case, the gapless phase is stable: it forms
stripes (as opposed to lines) in the phase diagram, bounded by $\mu_s=\pm 6t'\pm|\mu|$, as shown in
Fig.~\ref{fig:phasediag}. We have also computed the 
phase diagram numerically from Eq.~\eqref{eq:Dwmu}  for the values $t=4$, $\Delta=2$, $\mu=0.5$, which is also
shown in Fig.~\ref{fig:phasediag}; the numerical data agree with our analytically
obtained phase boundaries very well.

\section{Conclusions}
\label{sec:concl}
With the current interest in topological insulators and superconductors,
the role of the block offdiagonal structure [Eq.~\eqref{eq:offdiag}] in the literature
so far has been in the topological characterization of  gapped phases
of  ${\cal T}$-invariant superconductors. 
In this work we asked: what can be deduced from Eq.~\eqref{eq:offdiag} regarding
the nature of  gapless phases of these systems? Focusing on systems in $d=2,3$, 
our central finding is that Eq.~\eqref{eq:offdiag} leads to the possibility of $d-2$ dimensional
topologically stable Fermi surfaces, characterized by an integer topological charge $q$.
These Fermi surfaces can be envisioned as momentum space vortices in $\det D({\bf   k})$
as an "order parameter": in $d=2$, they are points (in $d=3$ they are lines) in momentum space where $\det D({\bf   k})$
vanishes and their topological charge is calculated by the phase winding of $\det D({\bf   k})$
around them. 

This result, combined  with the momentum reversal relations
in Eq.~\eqref{eqs:Drel}, leads to a "fermion quadrupling theorem": the
number of Fermi points for $d=2$ lattice models with a spin$\times$EH structure is a multiple of four. 
Our analogous finding for $d=3$ systems is that Fermi lines come in pairs.
The most well known manifestation of the fermion quadrupling theorem is provided by the four Fermi points of
a d-wave superconductor, and an example for a pair of Fermi lines is present in the noncentrosymmetric
superconductor CePt$_3$Si.\cite{Sato} 

In Sec.~\ref{sec:FSandq} we have seen that our results provide a straightforward topological
explanation for the robustness of the Fermi points of a d-wave superconductor against a weak 
spin-orbit coupling or against an admixture of a (time-reversal invariant) triplet pairing component.
In addition, our considerations regarding two dimensional lattice systems can find applications similar
to the work of Berg {\it et al}\cite{Berg2008} who studied the perturbative stability of Fermi points in the presence of
a (weak) coexisting order, with a periodicity commensurate with the lattice. For a ${\cal T}$-invariant
coexisting order,  the authors of Ref.~\onlinecite{Berg2008} found that the Fermi points cannot be gapped out as long as 
the perturbation does not couple them to each other. With our topological considerations at hand
we can say a little bit more:  the Fermi points are perturbatively unstable only if vortices are coupled to
antivortices. In this case, the resulting Fermi point in the folded Brillouin zone has  $q=0$, which guarantees
no topological stability; it can be gapped out by the perturbation. 
If Fermi points with like charges are coupled (for example a momentum reversed pair of Fermi points), 
the situation is different: in the folded Brillouin zone the resulting Fermi point has $|q|>1$. 
All what the  small perturbation can do is to split it up to more Fermi points.
Due to the quadrupling theorem, in addition, we know that the perturbation cannot remove only a single
vortex-antivortex pair; it has to reduce the number of Fermi points by four. 

We have also found several implications of our findings to topological superconductors. 
For the surface of a topological superconductor, we have related the net winding $q_{\rm net}$ to the bulk topological 
invariant $\nu$ in the form of an index theorem, Eq.~\eqref{eq:indexth}.
It accounts for the bulk topological properties in terms of a topological number for the 
boundary degrees of freedom, in perfect analogy to how the net number of chiral edge modes in a chiral
superconductor accounts for the Chern number\cite{Volovikbook} 
(the bulk topological invariant for chiral superconductors). 
Eq.~\eqref{eq:indexth} tells us that $\nu$ Fermi points on the surface are protected due to topological reasons,
but it also shows that topology does not forbid to have more Fermi points in the surface theory. 
These extra Fermi points, if present,  should be clearly visible, for example in the transport properties of the
topological superconductor surface.

Regarding the phase diagram for topological phase transitions, we have shown that generically,
because of the presence of topologically stable Fermi surfaces, there is an extended gapless region in parameter
space separating gapped phases with different $\nu$ -- gapless systems thus represent a separate,
stable phase themselves. 
This should be contrasted to the case where there can be no topologically stable Fermi surfaces. In this case, gapless
systems can only represent  (unstable) phase boundaries.

Finally, we note that our results related to topological superconductors can be extended to a bit 
broader context than the domains considered so far. First, 
as the symmetry classes in Eq.~\eqref{eqs:Drel} contain all the possibilities\cite{schnyderclass,schnyderAIP} 
in the Altland-Zirnbauer classification\cite{Alt97,heinzner2005symmetry}  for gapped three dimensional
 systems  with $\nu\in\mathbb{Z}$, our results can be seen as pertaining to 
"$\mathbb{Z}$-topological materials" in $d=3$, including chiral topological insulators\cite{schnyderclass,chZTI}
in class AIII in addition to superconductor systems. Second, as the key ingredient for
the appearance of a stable gapless phase across a topological phase transition is the existence
of topologically stable Fermi surfaces, 
such gapless phases are expected to appear also in two dimensional $\cal T$-invariant topological superconductors.
In these systems, the stability of the gapless phase is provided by the topological stability of Fermi points.

\ \\

\acknowledgments
The author has benefited from discussions with C.~W.~J. Beenakker
and A.~R. Akhmerov. This research was supported by the Dutch Science Foundation NWO/FOM.


\begin{thebibliography}{29}
\expandafter\ifx\csname natexlab\endcsname\relax\def\natexlab#1{#1}\fi
\expandafter\ifx\csname bibnamefont\endcsname\relax
  \def\bibnamefont#1{#1}\fi
\expandafter\ifx\csname bibfnamefont\endcsname\relax
  \def\bibfnamefont#1{#1}\fi
\expandafter\ifx\csname citenamefont\endcsname\relax
  \def\citenamefont#1{#1}\fi
\expandafter\ifx\csname url\endcsname\relax
  \def\url#1{\texttt{#1}}\fi
\expandafter\ifx\csname urlprefix\endcsname\relax\def\urlprefix{URL }\fi
\providecommand{\bibinfo}[2]{#2}
\providecommand{\eprint}[2][]{\url{#2}}

\bibitem[{\citenamefont{Schnyder et~al.}(2008)\citenamefont{Schnyder, Ryu,
  Furusaki, and Ludwig}}]{schnyderclass}
\bibinfo{author}{\bibfnamefont{A.~P.} \bibnamefont{Schnyder}},
  \bibinfo{author}{\bibfnamefont{S.}~\bibnamefont{Ryu}},
  \bibinfo{author}{\bibfnamefont{A.}~\bibnamefont{Furusaki}}, \bibnamefont{and}
  \bibinfo{author}{\bibfnamefont{A.~W.~W.} \bibnamefont{Ludwig}},
  \bibinfo{journal}{\prb} \textbf{\bibinfo{volume}{78}},
  \bibinfo{pages}{195125} (\bibinfo{year}{2008}).

\bibitem[{\citenamefont{Schnyder
  et~al.}(2009{\natexlab{a}})\citenamefont{Schnyder, Ryu, Furusaki, and
  Ludwig}}]{schnyderAIP}
\bibinfo{author}{\bibfnamefont{A.}~\bibnamefont{Schnyder}},
  \bibinfo{author}{\bibfnamefont{S.}~\bibnamefont{Ryu}},
  \bibinfo{author}{\bibfnamefont{A.}~\bibnamefont{Furusaki}}, \bibnamefont{and}
  \bibinfo{author}{\bibfnamefont{A.}~\bibnamefont{Ludwig}}, in
  \emph{\bibinfo{booktitle}{AIP Conf. Proc}}
  (\bibinfo{year}{2009}{\natexlab{a}}), vol. \bibinfo{volume}{1134},
  p.~\bibinfo{pages}{10}.

\bibitem[{\citenamefont{Kitaev}(2009)}]{kitaevtable}
\bibinfo{author}{\bibfnamefont{A.}~\bibnamefont{Kitaev}}, \bibinfo{journal}{AIP
  Conf. Proc.} \textbf{\bibinfo{volume}{1134}}, \bibinfo{pages}{22}
  (\bibinfo{year}{2009}).

\bibitem[{\citenamefont{Qi et~al.}(2009)\citenamefont{Qi, Hughes, Raghu, and
  Zhang}}]{qi:187001}
\bibinfo{author}{\bibfnamefont{X.-L.} \bibnamefont{Qi}},
  \bibinfo{author}{\bibfnamefont{T.~L.} \bibnamefont{Hughes}},
  \bibinfo{author}{\bibfnamefont{S.}~\bibnamefont{Raghu}}, \bibnamefont{and}
  \bibinfo{author}{\bibfnamefont{S.-C.} \bibnamefont{Zhang}},
  \bibinfo{journal}{\prl} \textbf{\bibinfo{volume}{102}},
  \bibinfo{pages}{187001} (\bibinfo{year}{2009}).

\bibitem[{\citenamefont{Roy}(2008)}]{roy2008topological}
\bibinfo{author}{\bibfnamefont{R.}~\bibnamefont{Roy}},
  \bibinfo{journal}{arXiv:0803.2868}  (\bibinfo{year}{2008}).

\bibitem[{\citenamefont{Schnyder
  et~al.}(2009{\natexlab{b}})\citenamefont{Schnyder, Ryu, and
  Ludwig}}]{schnydermodel}
\bibinfo{author}{\bibfnamefont{A.~P.} \bibnamefont{Schnyder}},
  \bibinfo{author}{\bibfnamefont{S.}~\bibnamefont{Ryu}}, \bibnamefont{and}
  \bibinfo{author}{\bibfnamefont{A.~W.~W.} \bibnamefont{Ludwig}},
  \bibinfo{journal}{\prl} \textbf{\bibinfo{volume}{102}},
  \bibinfo{pages}{196804} (\bibinfo{year}{2009}{\natexlab{b}}).

\bibitem[{\citenamefont{Ryu}(2009)}]{ryumodel}
\bibinfo{author}{\bibfnamefont{S.}~\bibnamefont{Ryu}}, \bibinfo{journal}{\prb}
  \textbf{\bibinfo{volume}{79}}, \bibinfo{pages}{075124}
  (\bibinfo{year}{2009}).

\bibitem[{\citenamefont{Hosur et~al.}(2009)\citenamefont{Hosur, Ryu, and
  Vishwanath}}]{chZTI}
\bibinfo{author}{\bibfnamefont{P.}~\bibnamefont{Hosur}},
  \bibinfo{author}{\bibfnamefont{S.}~\bibnamefont{Ryu}}, \bibnamefont{and}
  \bibinfo{author}{\bibfnamefont{A.}~\bibnamefont{Vishwanath}},
 \bibinfo{journal}{\prb}
  \textbf{\bibinfo{volume}{81}}, \bibinfo{pages}{045120}
  (\bibinfo{year}{2010}).


\bibitem[{\citenamefont{Salomaa and Volovik}(1988)}]{cosmic}
\bibinfo{author}{\bibfnamefont{M.~M.} \bibnamefont{Salomaa}} \bibnamefont{and}
  \bibinfo{author}{\bibfnamefont{G.~E.} \bibnamefont{Volovik}},
  \bibinfo{journal}{\prb} \textbf{\bibinfo{volume}{37}}, \bibinfo{pages}{9298}
  (\bibinfo{year}{1988}).

\bibitem[{\citenamefont{Volovik and Yakovenko}(1989)}]{volovik1989fractional}
\bibinfo{author}{\bibfnamefont{G.~E.} \bibnamefont{Volovik}} \bibnamefont{and}
  \bibinfo{author}{\bibfnamefont{V.~M.} \bibnamefont{Yakovenko}},
  \bibinfo{journal}{J. Phys.: Condens. Matter} \textbf{\bibinfo{volume}{1}},
  \bibinfo{pages}{5263} (\bibinfo{year}{1989}).

\bibitem[{\citenamefont{Volovik}(1992)}]{volovik1992exotic}
\bibinfo{author}{\bibfnamefont{G.~E.} \bibnamefont{Volovik}},
  \emph{\bibinfo{title}{{Exotic properties of superfluid $^3$He}}}
  (\bibinfo{publisher}{WS}, \bibinfo{year}{1992}).

\bibitem[{\citenamefont{Volovik}(2009)}]{volovik2009}
\bibinfo{author}{\bibfnamefont{G.~E.} \bibnamefont{Volovik}},
  \bibinfo{journal}{Pis'ma ZhETF} \textbf{\bibinfo{volume}{90}},
  \bibinfo{pages}{639} (\bibinfo{year}{2009}).

\bibitem[{TRE()}]{TREHfootnote}
\bibinfo{note}{Throughout this paper, deformations are always assumed to
  respect time-reversal and electron-hole symmetries.}

\bibitem[{\citenamefont{Volovik}(2003)}]{Volovikbook}
\bibinfo{author}{\bibfnamefont{G.~E.} \bibnamefont{Volovik}},
  \emph{\bibinfo{title}{{The Universe in a Helium droplet}}}
  (\bibinfo{publisher}{OUP, USA}, \bibinfo{year}{2003}).

\bibitem[{\citenamefont{Ho\ifmmode~\check{r}\else
  \v{r}\fi{}ava}(2005)}]{Horava}
\bibinfo{author}{\bibfnamefont{P.}~\bibnamefont{Ho\ifmmode~\check{r}\else
  \v{r}\fi{}ava}}, \bibinfo{journal}{Phys. Rev. Lett.}
  \textbf{\bibinfo{volume}{95}}, \bibinfo{pages}{016405}
  (\bibinfo{year}{2005}).

\bibitem[{\citenamefont{Volovik}(2007)}]{Volovikqpt}
\bibinfo{author}{\bibfnamefont{G.~E.} \bibnamefont{Volovik}},
  \bibinfo{journal}{Lect. Notes Phys.} \textbf{\bibinfo{volume}{718}},
  \bibinfo{pages}{31} (\bibinfo{year}{2007}).

\bibitem[{\citenamefont{Sato}(2006)}]{Sato}
\bibinfo{author}{\bibfnamefont{M.}~\bibnamefont{Sato}}, \bibinfo{journal}{\prb}
  \textbf{\bibinfo{volume}{73}}, \bibinfo{pages}{214502}
  (\bibinfo{year}{2006}).

\bibitem[{\citenamefont{Wen and Zee}(2002)}]{WenZee}
\bibinfo{author}{\bibfnamefont{X.~G.} \bibnamefont{Wen}} \bibnamefont{and}
  \bibinfo{author}{\bibfnamefont{A.}~\bibnamefont{Zee}},
  \bibinfo{journal}{Phys. Rev. B} \textbf{\bibinfo{volume}{66}},
  \bibinfo{pages}{235110} (\bibinfo{year}{2002}).

\bibitem[{ddi()}]{ddimnote}
\bibinfo{note}{For completeness, we note that Fermi surfaces of different dimensions would
  be possible if ${\cal C}'={\rm diag}(\openone_N,-\openone_{N'})$ ($N\neq N'$)
  was allowed. Such ${\cal C}'$ does not occur in the bulk of superconductors, neither
  on the surface of the topological superconductors considered here, therefore we focus solely on the
  $N=N'$ case.}

\bibitem[{\citenamefont{Ma{\~n}es et~al.}(2007)\citenamefont{Ma{\~n}es, Guinea,
  and Vozmediano}}]{Manes}
\bibinfo{author}{\bibfnamefont{J.~L.} \bibnamefont{Ma{\~n}es}},
  \bibinfo{author}{\bibfnamefont{F.}~\bibnamefont{Guinea}}, \bibnamefont{and}
  \bibinfo{author}{\bibfnamefont{M.~A.~H.} \bibnamefont{Vozmediano}},
  \bibinfo{journal}{Phys. Rev. B} \textbf{\bibinfo{volume}{75}},
  \bibinfo{pages}{155424} (\bibinfo{year}{2007}).

\bibitem[{loc()}]{localnote}
\bibinfo{note}{By local creation we mean that there exists a closed $d-1$
  dimensional surface enclosing the region where Fermi surfaces appear, or for lattice
  systems, that there is a suitably chosen reciprocal unit cell such that on
  its boundary $\det D(\veck)\neq 0$ throughout the deformation.}

\bibitem[{\citenamefont{Nielsen and Ninomiya}(1981)}]{NiNino}
\bibinfo{author}{\bibfnamefont{H.~B.} \bibnamefont{Nielsen}} \bibnamefont{and}
  \bibinfo{author}{\bibfnamefont{M.}~\bibnamefont{Ninomiya}},
  \bibinfo{journal}{Nucl. Phys. B} \textbf{\bibinfo{volume}{185}},
 \bibinfo{pages}{20} (\bibinfo{year}{1981}).

\bibitem[{\citenamefont{Hatsugai et~al.}(2004)\citenamefont{Hatsugai, Ryu, and
  Kohmoto}}]{Hatsugai}
\bibinfo{author}{\bibfnamefont{Y.}~\bibnamefont{Hatsugai}},
  \bibinfo{author}{\bibfnamefont{S.}~\bibnamefont{Ryu}}, \bibnamefont{and}
  \bibinfo{author}{\bibfnamefont{M.}~\bibnamefont{Kohmoto}},
  \bibinfo{journal}{Phys. Rev. B} \textbf{\bibinfo{volume}{70}},
  \bibinfo{pages}{054502} (\bibinfo{year}{2004}).

\bibitem[{\citenamefont{Altland and Zirnbauer}(1997)}]{Alt97}
\bibinfo{author}{\bibfnamefont{A.}~\bibnamefont{Altland}} \bibnamefont{and}
  \bibinfo{author}{\bibfnamefont{M.~R.} \bibnamefont{Zirnbauer}},
  \bibinfo{journal}{Phys.\ Rev.\ B} \textbf{\bibinfo{volume}{55}},
  \bibinfo{pages}{1142} (\bibinfo{year}{1997}).

\bibitem[{\citenamefont{Heinzner et~al.}(2005)\citenamefont{Heinzner,
  Huckleberry, and Zirnbauer}}]{heinzner2005symmetry}
\bibinfo{author}{\bibfnamefont{P.}~\bibnamefont{Heinzner}},
  \bibinfo{author}{\bibfnamefont{A.}~\bibnamefont{Huckleberry}},
  \bibnamefont{and}
  \bibinfo{author}{\bibfnamefont{M.}~\bibnamefont{Zirnbauer}},
  \bibinfo{journal}{Commun. Math. Phys.} \textbf{\bibinfo{volume}{257}},
  \bibinfo{pages}{725} (\bibinfo{year}{2005}).

\bibitem[{\citenamefont{Foster and Ludwig}(2008)}]{fosterludwig}
\bibinfo{author}{\bibfnamefont{M.~S.} \bibnamefont{Foster}} \bibnamefont{and}
  \bibinfo{author}{\bibfnamefont{A.~W.~W.} \bibnamefont{Ludwig}},
  \bibinfo{journal}{\prb} \textbf{\bibinfo{volume}{77}},
  \bibinfo{pages}{165108} (\bibinfo{year}{2008}).

\bibitem[{\citenamefont{Fu et~al.}(2007)\citenamefont{Fu, Kane, and
  Mele}}]{fukanemele}
\bibinfo{author}{\bibfnamefont{L.}~\bibnamefont{Fu}},
  \bibinfo{author}{\bibfnamefont{C.~L.} \bibnamefont{Kane}}, \bibnamefont{and}
  \bibinfo{author}{\bibfnamefont{E.~J.} \bibnamefont{Mele}},
  \bibinfo{journal}{\prl} \textbf{\bibinfo{volume}{98}},
  \bibinfo{pages}{106803} (\bibinfo{year}{2007}).

\bibitem[{\citenamefont{Murakami and ichi Kuga}(2008)}]{murakami}
\bibinfo{author}{\bibfnamefont{S.}~\bibnamefont{Murakami}} \bibnamefont{and}
  \bibinfo{author}{\bibfnamefont{S.}~\bibnamefont{ichi Kuga}},
  \bibinfo{journal}{\prb} \textbf{\bibinfo{volume}{78}},
  \bibinfo{pages}{165313} (\bibinfo{year}{2008}).

\bibitem[{\citenamefont{Berg et~al.}(2008)\citenamefont{Berg, Chen, and
  Kivelson}}]{Berg2008}
\bibinfo{author}{\bibfnamefont{E.}~\bibnamefont{Berg}},
  \bibinfo{author}{\bibfnamefont{C.-C.} \bibnamefont{Chen}}, \bibnamefont{and}
  \bibinfo{author}{\bibfnamefont{S.~A.} \bibnamefont{Kivelson}},
  \bibinfo{journal}{Phys. Rev. Lett.} \textbf{\bibinfo{volume}{100}},
  \bibinfo{pages}{027003} (\bibinfo{year}{2008}).

\end{thebibliography}
\end{document}